\documentclass[twocolumn,showpacs,preprintnumbers,amsmath,amssymb]{revtex4}

\usepackage{graphicx}
\usepackage{dcolumn}
\usepackage{bm}

\newcommand{\bq}{\begin{equation}}
\newcommand{\eq}{\end{equation}}
\newcommand{\bqa}{\begin{eqnarray}}
\newcommand{\eqa}{\end{eqnarray}}
\newcommand{\nn}{\nonumber \\}
\newcommand{\ij}{\langle i j \rangle}

\def\be     {\begin{equation}}
\def\ee     {\end{equation}}
\def\bea        {\begin{eqnarray}}
\def\eea        {\end{eqnarray}}
\def\bnn    {\begin{eqnarray*}}
\def\enn    {\end{eqnarray*}}

\begin{document}

\title{Heavy-fermion spin liquid in the strong hybridization limit of the finite-U Anderson lattice model}
\author{Ki-Seok Kim}
\affiliation{ School of Physics, Korea Institute for Advanced
Study, Seoul 130-012, Korea }
\date{\today}

\begin{abstract}
Studying the finite-U Anderson lattice model in the strong
hybridization limit, we find a heavy-fermion spin liquid phase,
where both conduction and localized fermions are strongly
hybridized to form heavy fermions but this heavy-fermion phase
corresponds to a symmetric Mott insulating state owing to the
presence of charge gap, resulting from large Hubbard-U
interactions in localized fermions. We show that this
heavy-fermion spin liquid phase differs from the "fractionalized"
Fermi liquid state, where the latter corresponds to a metallic
state with a small Fermi surface of conduction electrons while
localized fermions decouple from conduction electrons to form a
spin liquid state. We discuss the stability of this anomalous spin
liquid phase against antiferromagnetic ordering and gauge
fluctuations, in particular, instanton effects associated with
confinement of slave particles. Furthermore, we propose a
variational wave function to check its existence from the
microscopic model.
\end{abstract}


\maketitle

\section{Introduction}

Spin liquid has been intensively studied, motivated from the
resonating-valance-bond (RVB) scenario for anomalous finite
temperature physics of the Pseudogap phase in high $T_{c}$
cuprates.\cite{RVB_SL} However, such spin liquid physics is
fascinating itself because this phase has a nontrivial order
called topological or quantum order beyond the description of the
Landau-Ginzburg-Wilson paradigm for phase transitions, and its low
energy physics is described by a gauge theory, far from the
conventional structure of condensed matter theories, allowing
spin-fractionalized excitations called
spinons.\cite{Wen_Quantum_Order}

Recently, several geometrically frustrated insulators are proposed
to be genuine symmetric Mott insulating phases, i.e., spin
liquids.\cite{Spin_liquids} Although such magnetic frustration can
be a strong candidate as the mechanism for the existence of spin
liquid, it is still interesting to find another mechanism for its
existence. In this paper we propose a heavy-fermion spin liquid
phase in the strong hybridization limit of the finite-U Anderson
lattice model (ALM).

An important thing is how to suppress magnetic ordering.
Antiferromagnetic order is well known to occur at half filling in
the square lattice due to Fermi-nesting. One way to suppress the
antiferromagnetic ordering is to make spin singlets introducing
another band-electrons. Furthermore, if the nesting property can
be destroyed, such an ordering tendency will be much more
suppressed. This motivates us to consider the ALM since it shows a
large Fermi surface when conduction electrons are hybridized with
localized electrons.

In this paper we consider the strong hybridization limit, turning
on on-site Hubbard interactions to localized electrons. Two
possibilities are expected. One is that strong local interactions
weaken hybridization, causing the "fractionalized" Fermi liquid
phase, where localized fermions form a spin liquid phase while
conduction electrons are in the Fermi liquid
state.\cite{Senthil_Kondo} However, Fermi-nesting in the localized
fermion-band would result in antiferromagnetic ordering.
Magnetically frustrated interactions are required for the
fractionalized Fermi liquid to realize. The other is that the
hybridization still survives against local interactions, but
charge fluctuations are suppressed due to such interactions. Since
Fermi-nesting does not exist and Kondo singlets are formed owing
to hybridization, antiferromagnetic ordering does not arise, but
this corresponds to an insulating state owing to charge gap.
Because such a phase contains all symmetries of the original
lattice model, it can be identified with a spin liquid state with
heavy neutral fermions, i.e., spinons. We propose that this
heavy-fermion spin liquid state can emerge in the strong
hybridization limit of the finite-U ALM.

To describe Mott transition, it is necessary to take on-site
Hubbard interactions non-perturbatively. Recent studies on the
Hubbard model have revealed that the slave-rotor representation
can treat such interactions well, describing the Mott
transition.\cite{Kim_Rotor} Applying the U(1) slave-rotor
representation to the finite-U ALM, and performing gauge
transformation\cite{Kim_Kondo_SB} appropriate in the strong
hybridization limit, we find an effective U(1) gauge Lagrangian in
terms of renormalized conduction and localized fermions and
collective density-fluctuation bosons interacting via U(1)
slave-rotor gauge fluctuations. We show that such density
fluctuations associated with localized fermions become gapped,
increasing the Hubbard interaction-U in the strong hybridization
limit. This gapped phase is identified with an insulating phase
with all symmetries of the ALM, thus heavy-fermion spin liquid. We
discuss the stability of the spin liquid phase against gauge
fluctuations, in particular, instanton excitations associated with
confinement of slave-particles. Furthermore, we propose a
variational wave function to check its existence from not the
effective field theory but the microscopic model itself.

It should be noted that the heavy-fermion spin liquid phase
differs from the fractionalized Fermi liquid
state\cite{Senthil_Kondo} because the former is an insulator, thus
there is no Fermi surface while the latter is a metal to have a
small Fermi surface of conduction electrons. In this respect the
present Mott transition in the ALM should be discriminated from
Kondo breakdown as the orbital-selective Mott
transition\cite{Pepin}. Furthermore, the spin liquid state in the
fractionalized Fermi liquid phase would be unstable against
antiferromagnetic ordering owing to the presence of Fermi-nesting.

It is important to notice that the limit considered in this paper
is different from that studied in the context of heavy fermion
physics.\cite{Review_theory,Coleman_Review} Quantum phase
transitions in heavy fermion compounds are studied in the large-U
limit of the ALM, where the Kondo lattice model (KLM) or
infinite-U ALM is in the main interest. On the other hand, the
present study considers the strong hybridization limit, not
seriously taken into account in the KLM or infinite-U ALM context.

\section{U(1) slave-rotor representation of the finite-U Anderson lattice model}

We start from the finite-U ALM \bqa && L_{ALM} =
\sum_{i\sigma}c_{i\sigma}^{\dagger}(\partial_{\tau} -
\mu)c_{i\sigma} -
t\sum_{\ij\sigma}(c_{i\sigma}^{\dagger}c_{j\sigma} + H.c.) \nn &&
- V\sum_{i\sigma}(c_{i\sigma}^{\dagger}d_{i\sigma} + H.c.) +
\sum_{i\sigma}d_{i\sigma}^{\dagger}(\partial_{\tau} +
\epsilon_{d})d_{i\sigma} \nn && +
U\sum_{i}d_{i\uparrow}^{\dagger}d_{i\uparrow}d_{i\downarrow}^{\dagger}d_{i\downarrow}
, \eqa where $c_{i\sigma}$ represents conduction electrons with
dispersion $\epsilon_{k}^{c} = - 2t (\cos k_{x} + \cos k_{y})$ in
two dimensions, and $d_{i\sigma}$ localized electrons with
localized level $\epsilon_{d}$.

Quantum phase transitions are expected to occur via competition
between the hybridization term with $V$ and the local interaction
term with $U$. Without the interaction term of $U$ this model is
exactly solvable, resulting in hybridization between the
conduction and localized bands. Our problem is to study what
happens to the hybridization when the local interaction $U$ is
turned on.

The local interaction term can be decomposed into the charge and
spin channels \bqa &&
U\sum_{i}d_{i\uparrow}^{\dagger}d_{i\uparrow}d_{i\downarrow}^{\dagger}d_{i\downarrow}
=
\frac{U}{4}\sum_{i}(\sum_{\sigma}d_{i\sigma}^{\dagger}d_{i\sigma}
- 1)^{2} \nn && -
\frac{U}{4}\sum_{i}(\sum_{\sigma}\sigma{d}_{i\sigma}^{\dagger}d_{i\sigma})^{2}
+ \frac{U}{2}\sum_{i\sigma}d_{i\sigma}^{\dagger}d_{i\sigma} -
\frac{U}{4}\sum_{i}1 . \eqa In this paper we do not consider the
spin channel, justified in the strong hybridization limit where
antiferromagnetic ordering is severely suppressed.

Recently, we could describe the genuine Mott transition without
symmetry breaking, introducing the slave-rotor representation in
order to take into account local interactions non-perturbatively
in the Hubbard model.\cite{Kim_Rotor} Decomposing the localized
electron $d_{i\sigma} = e^{-i\theta_{i}}\eta_{i\sigma}$, one can
rewrite the ALM in the following way \bqa && L_{ALM} =
\sum_{i\sigma}c_{i\sigma}^{\dagger}(\partial_{\tau} -
\mu)c_{i\sigma} -
t\sum_{\ij\sigma}(c_{i\sigma}^{\dagger}c_{j\sigma} + H.c.) \nn &&
-
V\sum_{i\sigma}(c_{i\sigma}^{\dagger}e^{-i\theta_{i}}\eta_{i\sigma}
+ H.c.) + \sum_{i\sigma}\eta_{i\sigma}^{\dagger}(\partial_{\tau} +
\epsilon_{d})\eta_{i\sigma} \nn && + \frac{U}{4}\sum_{i}L_{i}^{2}
- i\sum_{i}L_{i}\partial_{\tau}\theta_{i} \nn && +
i\sum_{i}\varphi_{i}(L_{i} -
[\sum_{\sigma}\eta_{i\sigma}^{\dagger}\eta_{i\sigma} - 1]) , \eqa
where $\epsilon_{d}$ is replaced with $\epsilon_{d} - U/2$. It is
easy to show that Eq. (3) is exactly the same as Eq. (1) with Eq.
(2) after integrating out the $\varphi_{i}$ field with
$\eta_{i\sigma} = e^{i\theta_{i}}d_{i\sigma}$. In this expression
the electron Hilbert space $|d_{i\sigma}>$ is given by the direct
product of the fermion and boson Hilbert spaces
$|\eta_{i\sigma}>\bigotimes{|L_{i}>}$ according to the
decomposition $d_{i\sigma} = e^{-i\theta_{i}}\eta_{i\sigma}$,
where $L_{i}$ represents an electron density at site $i$. It is
clear that any decomposition method enlarges the original electron
Hilbert space, thus an appropriate constraint associated with the
decomposition should be imposed. The Lagrange multiplier field
$\varphi_{i}$ expresses the U(1) slave-rotor constraint $L_{i} =
\sum_{\sigma}\eta_{i\sigma}^{\dagger}\eta_{i\sigma} - 1$, implying
that the fermion and boson Hilbert spaces are not independent,
thus the two operators $\eta_{i\sigma}$ and $e^{i\theta_{i}}$
also. Then, $e^{-i\theta_{i}}$ is identified with an annihilation
operator of an electron charge owing to the constraint $L_{i} =
\sum_{\sigma}\eta_{i\sigma}^{\dagger}\eta_{i\sigma} - 1$ and the
canonical relation $[L_{i}, \theta_{j}] = -i\delta_{ij}$ imposed
by $- iL_{i}\partial_{\tau}\theta_{i}$. In this respect collective
density fluctuations associated with localized fermions can be
taken into account non-perturbatively in the U(1) slave-rotor
representation.

\section{Heavy fermion physics: Kondo breakdown as the orbital-selective Mott transition}

The problem is how to treat the hybridization term with
$e^{-i\theta_{i}}$. One direct way is to integrate out conduction
electrons and obtain an effective Lagrangian in terms of localized
fermions and their density-fluctuation bosons.\cite{Senthil_Kondo}
Integrating out the density field $L_{i}$ and conduction electron
field $c_{i\sigma}$, we obtain \bqa && L_{eff} =
\sum_{i\sigma}\eta_{i\tau\sigma}^{\dagger}(\partial_{\tau} +
\epsilon_{d})\eta_{i\tau\sigma} -
i\sum_{i}\varphi_{i\tau}[\sum_{\sigma}{\eta}^{\dagger}_{i\tau\sigma}{\eta}_{i\tau\sigma}
- 1] \nn && -
\sum_{ij\sigma}\eta_{i\tau\sigma}^{\dagger}e^{i\theta_{i\tau}}\Delta^{c}_{ij,\tau\tau'}e^{-i\theta_{j\tau'}}\eta_{j\tau'\sigma}
+ \frac{1}{U}\sum_{i}(\partial_{\tau}\theta_{i\tau} -
\varphi_{i\tau})^{2} , \nn \eqa where $\Delta^{c}_{ij,\tau\tau'}$
is the single particle propagator of the conduction electron,
given by \bqa && \Delta^{c}_{q,\omega} = \frac{V^{2}}{i\omega +
\mu - \epsilon_{q}^{c}} \nonumber  \eqa in the energy-momentum
space.

Although the above treatment itself is exact so far, it has an
important assumption that the identity of conduction electrons is
sustained. In other words, feedback effects of localized fermions
and density fluctuations to conduction electrons, self-energy
corrections of conduction electrons, are not introduced. Since
such feedback effects can be induced only via the hybridization
coupling term, this treatment may be regarded as the weak
hybridization approach, not justified in the strong hybridization
limit.\cite{HMM} Alternatively, it may be viewed as the
large-U-limit approach owing to small $V/U$, where the infinite-U
ALM can be considered or the KLM is taken via virtual charge
fluctuations. This type of treatment has been utilized both
intensively and extensively, associated with the
single-impurity\cite{FG} and heavy fermion
physics\cite{Coleman_Review}.

Performing the Hubbard-Stratonovich transformation for the
non-local hopping term in both time and space, we obtain an
effective Lagrangian \bqa && L_{eff} =
\sum_{i\sigma}\eta_{i\sigma}^{\dagger}(\partial_{\tau} +
\epsilon_{d})\eta_{i\sigma} -
i\sum_{i}\varphi_{i}[\sum_{\sigma}{\eta}^{\dagger}_{i\sigma}{\eta}_{i\sigma}
- 1] \nn && + \sum_{q} \chi_{q}^{\eta*}(\partial_{\tau} - \mu +
\epsilon_{q}^{c})\chi_{q}^{\theta} +
\frac{1}{U}\sum_{i}(\partial_{\tau}\theta_{i} - \varphi_{i})^{2}
\nn && -
V\sum_{ij}[\sum_{\sigma}\eta_{i\sigma}^{\dagger}\chi_{ij}^{\theta}\eta_{j\sigma}
+ e^{i\theta_{i}}\chi_{ij}^{\eta *}e^{-i\theta_{j}} ] , \eqa where
$\chi_{ij}^{\eta}$ and $\chi_{ij}^{\theta}$ are effective hopping
parameters determined in the self-consistent analysis \bqa &&
\chi_{ij}^{\eta*} = V \langle{e^{i\theta_{i}}(\partial_{\tau} -
\mu + \epsilon_{ij}^{c})^{-1}e^{-i\theta_{j}}}\rangle , \nn &&
\chi_{ij}^{\theta} = V
\langle{\eta_{i\sigma}^{\dagger}(\partial_{\tau} - \mu +
\epsilon_{ij}^{c})^{-1}\eta_{j\sigma}}\rangle . \eqa

It is not easy to perform the self-consistent analysis because
such effective hopping parameters have both frequency and momentum
dependencies (nonlocal in time and space). This non-locality may
give rise to crucial effects to the heavy fermion physics. Since
this is not our main subject, we leave its detailed analysis in an
important future problem, and here, discuss what is expected in
this treatment.

Generally speaking, two phases would be allowed in this
approximation, determined by the condensation of slave-rotor
bosons. In the large $V/U$ limit collective density-fluctuation
bosons become condensed, and heavy-fermion Fermi liquid appears to
form a large Fermi surface since $\langle{e^{i\theta_{i}}}\rangle
\not= 0$ results in
$\langle{c_{i\sigma}^{\dagger}\eta_{i\sigma}}\rangle \not= 0$ as
shown in the hybridization term of Eq. (3). In the small $V/U$
limit such boson excitations become gapped, and localized fermions
decouple from conduction electrons owing to
$\langle{c_{i\sigma}^{\dagger}\eta_{i\sigma}}\rangle = 0$, forming
a spin liquid state. Such a phase is called the fractionalized
Fermi liquid state with a small Fermi surface of conduction
electrons.\cite{Senthil_Kondo} As a result, the heavy-fermion
Fermi liquid to fractionalized Fermi liquid transition is
identified with Kondo breakdown as the orbital-selective Mott
transition\cite{Pepin} in the U(1) slave-rotor representation of
the finite-U ALM. However, the spin liquid state would be unstable
against antiferromagnetic ordering owing to Fermi-nesting in the
square lattice.

Although one important issue, that is, the abrupt volume-change of
the Fermi surface at the heavy-fermion quantum critical
point,\cite{YbRh2Si2} can be explained by the Kondo breakdown
transition as discussed above, the fact that the antiferromagnetic
transition of localized fermions arises simultaneously at the same
quantum critical point as the Kondo breakdown transition is still
far from our understanding because such a phenomenon is beyond the
description of the Landau-Ginzburg-Wilson theoretical framework
for phase transitions.\cite{LGW_Why} How to incorporate such an
antiferromagnetic transition in the Kondo breakdown one is an
important open problem.

\section{Strong hybridization approach}

In the previous section we have discussed the heavy fermion
physics in the "large-U" (more accurately, weak hybridization)
limit of the ALM although its mathematical construction is
performed at finite-U. In this section we take the strong
hybridization approach, considering that the hybridization term is
relevant in the renormalization group sense when $U = 0$. We show
that this type of approach allows a new possible phase called the
heavy-fermion spin liquid in the finite-U ALM.

\subsection{Gauge transformation and effective Lagrangian}

The relevance of the hybridization coupling in $U = 0$ motivates
us to consider the gauge transformation of $c_{i\sigma} =
e^{-i\theta_{i}}\psi_{i\sigma}$, where this type of approach has
been utilized in the context of quantum disordered
superconductors\cite{Quantum_disordered_SC}. Integrating out the
density field $L_{i}$ and inserting another slave-rotor
decomposition $c_{i\sigma} = e^{-i\theta_{i}}\psi_{i\sigma}$ into
Eq. (3), we find the following expression \bqa && Z =
\int{D\psi_{i\sigma}D\eta_{i\sigma}D\theta_{i}D\varphi_{i} }
\exp\Bigl[ - \int{d\tau} \Bigl\{
\sum_{i\sigma}\psi_{i\sigma}^{\dagger}(\partial_{\tau} \nn && -
\mu - i\partial_{\tau}\theta_{i})\psi_{i\sigma} -
t\sum_{\ij\sigma}(\psi_{i\sigma}^{\dagger}e^{i\theta_{i}}e^{-i\theta_{j}}\psi_{j\sigma}
+ H.c.)  \nn &&  - V\sum_{i\sigma}(\psi_{i\sigma}^{\dagger}
\eta_{i\sigma} + H.c.) +
\sum_{i\sigma}\eta_{i\sigma}^{\dagger}(\partial_{\tau} +
\epsilon_{d})\eta_{i\sigma} \nn && -
i\sum_{i}\varphi_{i}[\sum_{\sigma}{\eta}^{\dagger}_{i\sigma}{\eta}_{i\sigma}
- 1]  + \frac{1}{U}\sum_{i}(\partial_{\tau}\theta_{i} -
\varphi_{i})^{2} \Bigr\} \Bigr] . \eqa

Couplings between charge fluctuations and renormalized conduction
electrons can be decomposed using the Hubbard-Stratonovich
transformation. Following Ref. \cite{Kim_Rotor}, we find the
effective Lagrangian from Eq. (7) \bqa && L_{eff} = L_{0} +
L_{\psi} + L_{\eta} + L_{V} + L_{\theta} , \nn && L_{0} =
\frac{1}{U}\sum_{i}q_{ri}^{2} - \frac{2}{U}\sum_{i}q_{ri}
(\varphi_{ri} - p_{i}) + \sum_{\ij}txy , \nn && L_{\psi} =
\sum_{i\sigma}\psi_{i\sigma}^{\dagger}(\partial_{\tau} -
\mu)\psi_{i\sigma}  -
i\sum_{i}p_{i}[\sum_{\sigma}\psi_{i\sigma}^{\dagger}
\psi_{i\sigma} - 1] \nn && -
tx\sum_{\ij\sigma}(\psi_{i\sigma}^{\dagger}e^{-ia_{ij}}\psi_{j\sigma}
+ H.c.) , \nn && L_{\eta} =
\sum_{i\sigma}\eta_{i\sigma}^{\dagger}(\partial_{\tau} +
\epsilon_{d})\eta_{i\sigma} -
i\sum_{i}(\varphi_{ri}-q_{ri})[\sum_{\sigma}{\eta}^{\dagger}_{i\sigma}{\eta}_{i\sigma}
- 1] ,  \nn && L_{V} = - V\sum_{i\sigma}(\psi_{i\sigma}^{\dagger}
\eta_{i\sigma} + H.c.) , \nn && L_{\theta} =
\frac{1}{U}\sum_{i}(\partial_{\tau}\theta_{i} - \varphi_{ri})^{2}
- 2ty\sum_{\ij}\cos(\theta_{i}-\theta_{j}+a_{ij}) , \eqa where the
hopping parameters are represented as $x_{ij} = xe^{ia_{ij}}$ and
$y_{ij} = y e^{ia_{ij}}$, and $\varphi_{i} = \varphi_{ri} -
q_{ri}$ is used with $q_{ri} = {Uq_{i}}/{2}$. The unidentified
parameters can be determined self-consistently in the saddle-point
analysis \bqa && p_{i} = \langle
\partial_{\tau}\theta_{i}\rangle , ~~~~~  q_{ri} =
i\frac{U}{2}\langle\sum_{\sigma}\psi_{i\sigma}^{\dagger}\psi_{i\sigma}
- 1\rangle ,  \nn && x = |\langle e^{-i\theta_{i}}e^{i\theta_{j}}
+ H.c.\rangle| , ~~~~~  y =
|\langle\sum_{\sigma}\psi_{i\sigma}^{\dagger}\psi_{j\sigma} +
H.c.\rangle| . \nn \eqa $p_{i}$ and $q_{ri}$ play the role of
renormalization for the fermion and boson chemical potentials,
respectively.

The effective Lagrangian Eq. (8) is the main result of this paper,
showing an important feature of the finite-U ALM, that is, the
presence of two kinds of relevant interactions. As mentioned
before, the hybridization coupling is relevant to give rise to a
band-hybridized metal if the local charge-fluctuation energy is
not taken into account. On the other hand, an appropriate
treatment of local charge fluctuations has been shown to result in
the Mott transition from a spin liquid Mott insulator to a Fermi
liquid metal at a critical value $U_{c}$ without hybridization,
i.e., in the Hubbard model.\cite{Kim_Rotor}

Compared to the effective Lagrangian Eq. (5) for the heavy-fermion
quantum phase transition, the effective Lagrangian Eq. (8) always
allows the band-hybridization while Eq. (5) does not. In Eq. (5)
the boson condensation corresponds to the hybridization transition
since it gives an effective hybridization coupling constant
$V_{eff} = |\langle{e^{-i\theta_{i}}}\rangle|V$. Such a transition
is involved with the abrupt volume change of the Fermi surface. On
the other hand, in Eq. (8) the renormalized conduction and
localized fermions are always hybridized, forming a large Fermi
surface of spinons (neutral fermions). Then, the boson
condensation in Eq. (8) corresponds to the genuine Mott transition
in these heavy fermions.

Since the slave-rotor variable has its own dynamics in the
saddle-point approximation, given by the U(1) rotor (XY) model,
the quantum transition described by condensation of the rotor
field occurs when $Dy/U \sim 1$ with the half bandwidth $D$, where
$Dy$ is an effective bandwidth for the rotor
variable.\cite{Rotor_model} It is important to understand that the
ratio $V/U$ between the hybridization and on-site interactions
controls the hopping parameter $y$, where the hopping parameter
decreases as this ratio decreases. This results in the quantum
critical point $(V/U)_{c}$ between the heavy-fermion Fermi liquid
and heavy-fermion spin liquid, which differs from the
heavy-fermion quantum critical point between the heavy-fermion and
fractionalized Fermi liquids.

It is interesting to notice that this treatment [Eq. (8)] gives
rise to two different energy scales.\cite{Two_scales} One
corresponds to the band-hybridization scale $T_K$, and the other
is associated with the coherence scale $T_{FL}$, resulting in
Fermi liquid. $T_{K}$ is proportional to the hybridization gap $V$
in the strong coupling approach, consistent with the single
impurity energy scale in the strong coupling limit. On the other
hand, $T_{FL}$ is proportional to the effective stiffness
parameter $Dy|\langle{e^{i\theta_{i}}}\rangle|^{2}$ controlling
the phase coherence of $\theta_{i}$.\cite{TK_TFL}

\subsection{Mean-field analysis}

For the saddle-point analysis we resort to large $N$
generalization replacing $e^{i\theta_{i}}$ with $\phi_{is}$, where
$s = 1, ..., N$.\cite{Kim_Kim} Taking the mean-field ansatz of
$ip_{i} = p$, $i\varphi_{ri} = \varphi_{r}$, $iq_{ri} = - q_{r}$,
and $i \lambda_{i} = \lambda$ where $\lambda_{i}$ introduces the
rotor constraint $\sum_{s}|\phi_{is}|^{2} = 1$, we obtain the free
energy functional \bqa && F_{MF} = \nn &&  -
\frac{1}{\beta}\sum_{k\sigma}\sum_{\omega_{n}}\ln(i\omega_{n} -
E_{k+}) -
\frac{1}{\beta}\sum_{k\sigma}\sum_{\omega_{n}}\ln(i\omega_{n} -
E_{k-}) \nn && + \frac{1}{\beta}\sum_{ks}\sum_{\nu_n}\ln\Bigl( -
\frac{1}{U}[i\nu_{n} + \varphi_{r}]^{2} + y\epsilon_{k}^{\phi} +
\lambda \Bigr)  \nn && + \sum_{k}\Bigl(Dxy - \frac{2}{U}
q_{r}[\varphi_{r} - p + \frac{q_{r}}{2}] + p + \varphi_{r} + q_{r}
\nn && - \lambda + \mu [1-\delta]\Bigr) . \eqa Here the
renormalized fermion spectrum is given by $E_{k\pm} =
[\frac{(x\epsilon_{k}^{\psi}-\mu - p ) + (\epsilon_{d} -
\varphi_{r} - q_{r})}{2}] \pm
\sqrt{[\frac{(x\epsilon_{k}^{\psi}-\mu - p ) - (\epsilon_{d} -
\varphi_{r} - q_{r})}{2}]^{2} + V^{2}}$, and $\delta$ is hole
concentration for the conduction band. $\omega_{n}$ ($\nu_{n}$) is
the Matzubara frequency for fermions (bosons).

Minimizing the free energy Eq. (10) with respect to $\lambda$,
$x$, $y$, $q_{r}$, $\varphi_{r}$, $p$, and $\mu$, we obtain the
self-consistent mean-field equations. Performing the Matzubara
frequency summations and momentum integrals with $\sum_{k} =
\frac{1}{2D}\int_{-D}^{D}d\epsilon$ (constant density of
states\cite{Kim_Rotor,Kim_Kim}), we find \bqa && 1 =
\frac{\sqrt{U(\lambda+Dy)} -
\sqrt{U(\lambda+y\epsilon_{\theta})}}{Dy} ,  ~~~~~
\epsilon_{\theta} = \frac{1}{y}\Bigl(\frac{\varphi_{r}^{2}}{U} -
\lambda\Bigr) , \nn && x = \frac{(2\lambda - Dy)\sqrt{U(\lambda +
Dy)} - (2\lambda -
y\epsilon_{\theta})\sqrt{U(\lambda+y\epsilon_{\theta})}}{3(Dy)^2}
, \nn && y = \frac{1}{4}\Bigl(1 - \frac{\epsilon_{\psi}^2}{D^2}
\Bigr) + \frac{V^2}{2D^2x^2} \Bigl[ \ln \Bigl\{
\frac{\sin[\tan^{-1}(\frac{x\epsilon_{\psi}-\mu_r}{2V})]-1}{-\sin[\tan^{-1}(\frac{Dx+\mu_r}{2V}
)]-1}\Bigr\} \nn && -
\Bigl\{\frac{1}{\sin[\tan^{-1}(\frac{x\epsilon_{\psi}-\mu_r}{2V})]-1}
- \frac{1}{-\sin[\tan^{-1}(\frac{Dx+\mu_r}{2V} )]-1} \Bigr\} \nn
&& - \ln \Bigl\{
\frac{\sin[\tan^{-1}(\frac{x\epsilon_{\psi}-\mu_r}{2V})]+1}{-\sin[\tan^{-1}(\frac{Dx+\mu_r}{2V}
)]+1}\Bigr\} \nn && -
\Bigl\{\frac{1}{\sin[\tan^{-1}(\frac{x\epsilon_{\psi}-\mu_r}{2V})]+1}
- \frac{1}{-\sin[\tan^{-1}(\frac{Dx+\mu_r}{2V} )]+1} \Bigr\}
\Bigr] \nn && +
\frac{\mu_rV}{D^2x^2}\Bigl[\frac{1}{\cos[\tan^{-1}(\frac{x\epsilon_{\psi}-\mu_r}{2V})]}
- \frac{1}{\cos[\tan^{-1}(\frac{Dx+\mu_r}{2V} )]} \Bigr] , \nn &&
\epsilon_{\psi} = \frac{1}{x}\Bigl( \frac{V^2}{\epsilon_{d} -
q_{r}-\varphi_{r}} + \mu + p \Bigr) , \nn &&
\frac{2}{U}(\varphi_{r} - p + q_r) - 1 =  - \frac{1}{2}\Bigl(1 +
\frac{\epsilon_{\psi}}{D}\Bigr) \nn && -
\frac{V}{Dx}\Bigl[\frac{1}{\cos[\tan^{-1}(\frac{x\epsilon_{\psi}-\mu_r}{2V})]}
- \frac{1}{\cos[\tan^{-1}(\frac{Dx+\mu_r}{2V} )]} \Bigr] ,  \nn &&
\frac{2}{U}(\varphi_{r} - p) = (1 + \frac{\epsilon_{\theta}}{D}) =
( 1 - \frac{\epsilon_{\psi}}{D}) , ~~~~~ q_{r} = - \frac{U}{2}
\delta , \nn &&  \mu_{r} = \mu + p + \epsilon_{d} - q_r -
\varphi_r . \eqa

Condensation of the density-fluctuation bosons occurs when their
excitation gap closes or its effective chemical potential
vanishes, i.e., $\varphi_{r} = 0$. This causes $p = 0$, resulting
in $\epsilon_{\theta} = - D$ ($\epsilon_{\psi} = D$). We obtain
$\lambda = Dy$ at the quantum critical point, thus find the
relation $Dy/U = 2$, completely consistent with the previous
qualitative analysis for the rotor model.\cite{Rotor_model} The
hopping parameter for renormalized conduction fermions is $x =
1/3$ at the quantum critical point. As a result, the quantum
critical point $(V/D, U/D)_{c}$ is determined by \bqa &&
\frac{2U}{D} = \frac{V^2}{2(D/3)^2} \Bigl[ \ln \Bigl\{
\frac{\sin[\tan^{-1}(\frac{D/3-\mu_r}{2V})]-1}{-\sin[\tan^{-1}(\frac{D/3+\mu_r}{2V}
)]-1}\Bigr\}  \nn && -
\Bigl\{\frac{1}{\sin[\tan^{-1}(\frac{D/3-\mu_r}{2V})]-1} -
\frac{1}{-\sin[\tan^{-1}(\frac{D/3+\mu_r}{2V} )]-1} \Bigr\} \nn &&
- \ln \Bigl\{
\frac{\sin[\tan^{-1}(\frac{D/3-\mu_r}{2V})]+1}{-\sin[\tan^{-1}(\frac{D/3+\mu_r}{2V}
)]+1}\Bigr\}  \nn && -
\Bigl\{\frac{1}{\sin[\tan^{-1}(\frac{D/3-\mu_r}{2V})]+1} -
\frac{1}{-\sin[\tan^{-1}(\frac{D/3+\mu_r}{2V} )]+1} \Bigr\} \Bigr]
\nn && +
\frac{\mu_rV}{(D/3)^2}\Bigl[\frac{1}{\cos[\tan^{-1}(\frac{D/3-\mu_r}{2V})]}
- \frac{1}{\cos[\tan^{-1}(\frac{D/3+\mu_r}{2V} )]} \Bigr] , \nn &&
\delta =
\frac{V}{D/3}\Bigl[\frac{1}{\cos[\tan^{-1}(\frac{D/3-\mu_r}{2V})]}
- \frac{1}{\cos[\tan^{-1}(\frac{D/3+\mu_r}{2V} )]} \Bigr] , \nn &&
\mu_r = \frac{D}{3} - \Bigl( \frac{V^2}{\epsilon_{d} +
\frac{U}{2}\delta } - [\epsilon_{d} + \frac{U}{2}\delta ] \Bigr) .
\eqa Actually, we find $(0.112,0.044)$ at $\delta = 0.800$ and
$\epsilon_{d}/D = - 0.001$, thus $(V/U)_{c} = 2.545$.

This quantum phase transition results from gapping of density
fluctuations of localized fermions in the presence of strong
hybridization with renormalized conduction fermions, thus
differing from the Kondo breakdown transition\cite{Senthil_Kondo}
in the ALM. In $V/U > (V/U)_{c}$ collective density fluctuations
become softened, and the saddle-point equation of $\lambda$ is
replaced with $1 = \sqrt{2U/Dy} + Z$, where $Z =
|\langle{e^{i\theta_{i}}}\rangle|^{2}$ is the condensation
amplitude. Thus, the coherence temperature is given by $T_{FL}
\approx (Dy - 2U)^{1/2}$ near the heavy-fermion Mott critical
point, and below this temperature the valance-fluctuation-induced
heavy-fermion phase arises.

\section{Heavy-fermion spin liquid to heavy-fermion Fermi liquid quantum transition: its critical field theory and non-Fermi liquid physics}

To investigate low energy physics near the heavy-fermion spin
liquid to Fermi liquid quantum critical point, we derive an
effective field theory from Eq. (8). From the fermion mean-field
Lagrangian $L_{\chi} =
\sum_{k\sigma}\Bigl[\chi_{k\sigma+}(\partial_{\tau} +
E_{k+})\chi_{k\sigma+} + \chi_{k\sigma-}(\partial_{\tau} +
E_{k-})\chi_{k\sigma-}\Bigr]$ where $\psi_{k\sigma} =
u_{k}\chi_{k\sigma+} + v_{k}\chi_{k\sigma-}$ and $\eta_{k\sigma} =
v_{k}\chi_{k\sigma+} - u_{k}\chi_{k\sigma-}$ with $u_{k} =
{V}/{\sqrt{(E_{k+} - \epsilon_{k}^{\psi})^{2} + V^{2}}}$ and
$v_{k} = - ({E_{k+} - \epsilon_{k}^{\psi}})/{\sqrt{(E_{k+} -
\epsilon_{k}^{\psi})^{2} + V^{2}}}$, we find its critical field
theory ${\cal L}_{\chi} = \chi_{r\sigma}^{\dagger}(\partial_{\tau}
- \mu_{c})\chi_{r\sigma} + \frac{1}{2m_{\chi}}|(\nabla -
i\mathbf{a}_{r})\chi_{r\sigma}|^{2}$. Here the renormalized
fermion field $\chi_{r\sigma}$ represents $\chi_{i\sigma-}$ in the
continuum limit, and the effective band mass is given by $m_{\chi}
= \Bigl({\partial^{2}E_{k-}}/{\partial k^2}\Bigr)^{-1} =
{m_{e}}/\Bigl[{\frac{\epsilon_{dr}}{\epsilon_{dr}^2 +
V^2}\Bigl(\mu_{c} + \frac{V^2}{\epsilon_{dr}}\Bigr) - 1}\Bigr]$
with the "bare" band mass $m_{e} \sim (tx)^{-1}$, renormalized
localized level $\epsilon_{dr} = \epsilon_{d} +
\frac{U}{2}\delta$, and critical chemical potential $\mu_{c}$.
Note that $m_{\chi}$ diverges in the limit of $V \rightarrow
\infty$.

The effective boson Lagrangian can be written with an
electromagnetic field $\mathbf{A}$ as ${\cal L}_{\phi} =
[(\partial_{\tau} +
i\varphi_{r})\phi_{rs}^{\dagger}][(\partial_{\tau} -
i\varphi_{r})\phi_{rs}] + |(\nabla - i\mathbf{a}_{r} -
i\mathbf{A})\phi_{rs}|^{2} + m_{\phi}^{2}|\phi_{rs}|^{2} +
\frac{u_{\phi}}{2}|\phi_{rs}|^{2}$, where the rotor field
$e^{i\theta_{r}}$ is replaced with $\phi_{rs}$, and the rotor
constraint is softened via introduction of local interactions
$u_{\phi}$. $m_{\phi}$ is an effective mass, given by
$m_{\phi}^{2} = (V/U)_{c} - (V/U)$. An important point is that the
renormalized boson chemical potential $\varphi_{r}$ becomes zero
at the quantum critical point ($m_{\phi}^{2} = 0$), thus the
linear time-derivative term vanishes.

We find the critical field theory at the heavy-fermion spin liquid
quantum critical point \bqa && S_{c} = \int{d\tau}d^2x \Bigl[
\chi_{r\sigma}^{\dagger}(\partial_{\tau} - \mu_{c})\chi_{r\sigma}
+ \frac{1}{2m_{\chi}}|(\nabla -
i\mathbf{a}_{r})\chi_{r\sigma}|^{2} \nn && +
|\partial_{\tau}\phi_{rs}|^{2} + |(\nabla - i\mathbf{a}_{r} -
i\mathbf{A})\phi_{rs}|^{2}  + \frac{u_{\phi}}{2}|\phi_{rs}|^{2}
\Bigr] \nn && + S_{eff}[\mathbf{a}_r] , \eqa where the critical
gauge action is given by $S_{eff}[\mathbf{a}_{r}] =
\frac{1}{2}\sum_{q,\omega_{n}}\Bigl(
\gamma_{F}\frac{|\omega_{n}|}{q} + \frac{N}{8}q\Bigr)
\Bigl(\delta_{ij} - \frac{q_{i}q_{j}}{q^2}\Bigr)a_{ir}a_{jr}$ with
the damping strength $\gamma_{F} = k_{F}/\pi$ and boson flavor
number $N$.\cite{Kim_Kim} The first term represents dissipative
dynamics in gauge fluctuations due to particle-hole excitations of
$\chi_{r\sigma}$ fermions near the Fermi surface, and the second
term arises from critical boson fluctuations at the quantum
critical point.\cite{Kim_Kim} The Maxwell gauge action is omitted
since it is irrelevant in the renormalization group sense. Note
that the time component of the gauge field mediates a local
interaction, thus safely ignored in the low energy
limit.\cite{Sigma}

In the random phase approximation one finds the following
expression for the free energy $F/V = \int\frac{d^2q}{(2\pi)^{2}}
\int\frac{d\omega}{2\pi}\coth\Bigl[\frac{\omega}{2T}\Bigr]
\tan^{-1}\Bigl[\frac{{\rm Im}D(q,\omega)}{{\rm
Re}D(q,\omega)}\Bigr]$, where the gauge kernel is given by
$D(q,\omega) = \Bigl( - i \gamma_{F}\frac{\omega}{q} +
\frac{N}{8}q\Bigr)^{-1}$ in real frequency at the quantum critical
point. This free energy leads to the divergent specific heat
coefficient $\gamma = {C}_{V}/T \propto - \ln T$. The dc
conductivity is given by the Ioffe-Larkin combination-rule
$\sigma_{tot} = {\sigma_{\phi}\sigma_{\chi}}/({\sigma_{\phi}+
\sigma_{\chi}})$, assuming gaussian gauge
fluctuations.\cite{IL_sigma} In the one loop diagram of
gauge-boson exchange one finds ${1}/{\tau_{tr}^{\chi}} \sim
T^2$\cite{Kim_Kim} and $1/{\tau_{tr}^{\phi}} \sim T$\cite{Sigma},
where $1/{\tau_{tr}^{\chi}}$ ($1/{\tau_{tr}^{\phi}}$) is the
scattering rate of the $\chi_{r\sigma}$ fermion ($\phi_{rs}$
boson) with temperature $T$. As a result, the total dc
conductivity is given by $\sigma_{tot} \sim T^{-1}$ in low
temperatures, giving rise to the linear resistivity $\rho_{tot}
\sim T$ at the quantum critical point. On the other hand, in $V/U
< (V/U)_{c}$ but $|V/U - (V/U)_{c}| \rightarrow 0$ corresponding
to the heavy-fermion spin liquid state in the zero temperature
limit, the gauge kernel shows the crossover behavior from
$D(q,\omega) = \Bigl( - i \gamma_{F}\frac{\omega}{q} +
\frac{N}{8}q\Bigr)^{-1}$ to $D(q,\omega) = \Bigl( - i
\gamma_{F}\frac{\omega}{q} + \frac{q^2}{g}\Bigr)^{-1}$ with an
effective internal gauge-charge $g$ as temperature goes down. Here
the crossover energy scale would be an excitation gap of
$\phi_{rs}$, given by $|\varphi_{r}|$. Accordingly, the specific
heat coefficient exhibits the upturn behavior from $\gamma \sim -
\ln T$ to $\gamma \sim T^{-1/3}$. But, the resistivity still
behaves as $\rho_{tot} \sim T$ owing to $\rho_{\chi} \sim T^{4/3}$
in the $z = 3$ critical field theory,\cite{Sigma} where $z$ is the
dynamical exponent.

\section{Discussion and summary}

\subsection{Stability of the heavy-fermion spin liquid against gauge fluctuations}

The stability of the heavy-fermion spin liquid state against gauge
fluctuations is an important problem for this phase to realize in
the finite-U ALM. It has been shown that the fermion-gauge action
in Eq. (13), obtained via integrating out gapped boson
excitations, has an infrared stable interacting fixed point with
an effective nonzero internal charge when the flavor number or
density of fermions is sufficiently large.\cite{IR_FP} Actually,
the present author has studied that such a fixed point can arise
when the fermion conductivity, given by its current-current
correlation function, is sufficiently large.\cite{Kim_SL_FS}
Notice that the fermion conductivity is associated with the
density of fermions. At this fixed point the critical field theory
is characterized by the dynamical exponent $z = 3$, as discussed
before.

The problem is the stability of this fixed point against instanton
excitations resulting from compact gauge fluctuations. A similar
situation appears in the compact QED$_{3}$, where Dirac fermions
interact via compact U(1) gauge fluctuations. The infrared
interacting fixed point in the QED$_{3}$ has been shown to be
stable against instanton fluctuations since the scaling dimension
of an instanton insertion operator is proportional to the flavor
number of massless Dirac fermions, thus irrelevant in the large
flavor limit.\cite{Hermele} Following the similar strategy with
this case, the present author has shown that the scaling dimension
of the instanton operator is proportional to the fermion
conductivity analogous to the flavor number of Dirac fermions,
thus irrelevant in the large conductivity limit.\cite{Kim_SL_FS}
This seems to be natural because the fermion conductivity is
associated with screening of gauge interactions.

In the heavy-fermion spin liquid phase the fermion conductivity
may not be sufficiently large because the strong hybridization
makes the hybridized band flat. Intuitively, the existence of the
heavy-fermion spin liquid phase is not clear because the strong
hybridization coupling is necessary for its existence, but such
hybridization prohibits this phase from being stable against gauge
fluctuations. Since its stability depends on the fermion
conductivity,\cite{Kim_SL_FS,Nagaosa} to determine its presence
with gauge interactions is beyond the scope of the present study.
However, its existence is allowed in principle.

The stability of this phase against antiferromagnetic ordering is
confirmed since the Fermi-surface nesting does not exist in the
heavy-fermion band. This not only justifies our neglect of spin
fluctuations, but also propose new mechanism for the existence of
the spin liquid phase.

\subsection{A variational wave function}

The above discussion motivates us to find the heavy-fermion spin
liquid phase from the microscopic model itself, here the finite-U
ALM. In this respect we propose a variational wave function for
its existence. Since we start from the conduction-valance
hybridized state, the following wave function is naturally
proposed \bqa && |\Phi\rangle =
\Pi_{k\sigma}(c_{k\sigma}^{\dagger} +
a(k)d_{k\sigma}^{\dagger})|0\rangle , \eqa where $|0\rangle$ is
the vacuum state and $a(k)$ is the variational parameter given by
$a(k) = V/[(\epsilon_{k}^{c} - \mu - \epsilon_{d})/2 +
\sqrt{(\epsilon_{k}^{c} - \mu - \epsilon_{d})^{2}/4 + V^{2}}]$ in
$U = 0$. Turning on on-site interactions to localized electrons,
the variational ground state is proposed to
be\cite{Valance_HF_GPGW} \bqa |\Psi> && = P_{PG}|\Phi\rangle \nn
&& = \Pi_{i}(1 - \kappa n_{i\uparrow}^{d}n_{i\downarrow}^{d})
\Pi_{k\sigma}(c_{k\sigma}^{\dagger} +
a(k)d_{k\sigma}^{\dagger})|0> , \eqa where $P_{PG} = \Pi_{i}(1 -
\kappa n_{i\uparrow}^{d}n_{i\downarrow}^{d})$ is the partial
Gutzwiller projector, suppressing charge-fluctuation effects, with
an interaction-dependent $\kappa$ determined self-consistently as
a function of $U$.\cite{Gassamer_SC}

Since this ground-state wave function captures both relevant
interactions, $V$ and $U$ appropriately, physics included in Eq.
(15) is expected to be essentially imposed in the slave-rotor
effective gauge theory [Eq. (8)]. When gauge fluctuations are
ignored as the saddle-point approximation in Eq. (8), the
$|\Phi\rangle$ state is exactly recovered from the fermion
Lagrangian $L_{F} = L_{\psi} + L_{\eta} + L_{V}$ because $L_{F}$
gives rise to the same $a(k)$ in $|\Phi\rangle$. Note that the
gauge transformation $c_{i\sigma} =
e^{-i\theta_{i}}\psi_{i\sigma}$ is an essential procedure for
obtaining this strong hybridized dynamics. On the other hand, the
partial Gutzwiller projection operator $P_{PG}$ is simulated in
the slave-rotor representation, $L_{\theta}$ taking into account
on-site interactions of localized electrons appropriately. This
approach is parallel to that in the doped Mott insulator problem,
where the Gutzwiller projected BCS wave function, i.e., the RVB
state had been proposed,\cite{RVB} and such a variational ground
state was simulated in the context of the gauge
theory.\cite{RVB_SL} It will be an interesting project to check
whether this ground state wave function can result in both the
fractionalized Fermi liquid and heavy-fermion spin liquid.

\subsection{Summary}

In this paper we have studied the finite-U ALM in the strong
hybridization limit. The conventional treatment, integrating out
conduction electrons to obtain an effective Lagrangian in terms of
localized fermions and collective density-fluctuation bosons,
describes the heavy-fermion quantum transition from the
heavy-fermion Fermi liquid with a large Fermi surface to the
fractionalized Fermi liquid with a small Fermi surface, where the
abrupt volume change of the Fermi surface is involved. On the
other hand, the strong hybridization approach shows the
heavy-fermion spin liquid to heavy-fermion Fermi liquid
transition, where this type of Mott transition differs from the
orbital-selective one via Kondo breakdown.

We propose a schematic phase diagram of the finite-U ALM in Fig.
1, where HF-FL, HF-SL, and F-SL represent the heavy-fermion Fermi
liquid, heavy-fermion spin liquid, and fractionalized Fermi
liquid, respectively. It is important to notice that this model
has two independent parameters scaled by the conduction bandwidth,
$(V/D, U/D)$. In the small $U/D$ limit the hybridization coupling
is relevant to form a hybridized band, thus the heavy-fermion
Fermi liquid arises. In the large $U/D$ limit the Kondo breakdown
transition has been shown to occur in the KLM or infinite-U ALM,
thus there is a critical $V/D$ separating the fractionalized Fermi
liquid ($\langle{c_{\sigma}^{\dagger}\eta_{\sigma}}\rangle = 0$)
from the heavy-fermion Fermi liquid
($\langle{c_{\sigma}^{\dagger}\eta_{\sigma}}\rangle \not= 0$). On
the other hand, in the large $V/D$ limit the heavy-fermion band is
formed first
($\langle{\psi_{\sigma}^{\dagger}\eta_{\sigma}}\rangle \not= 0$),
and increasing $U/D$ (marked by the arrow line) is expected to
cause the Mott transition ($\langle{e^{-i\theta}}\rangle = 0$
$\rightarrow$ $\langle{c_{\sigma}^{\dagger}\eta_{\sigma}}\rangle
\approx \langle{e^{-i\theta}}\rangle
\langle{\psi_{\sigma}^{\dagger}\eta_{\sigma}}\rangle = 0$) in this
heavy-fermion band.

\begin{figure}
\includegraphics[width=8cm]{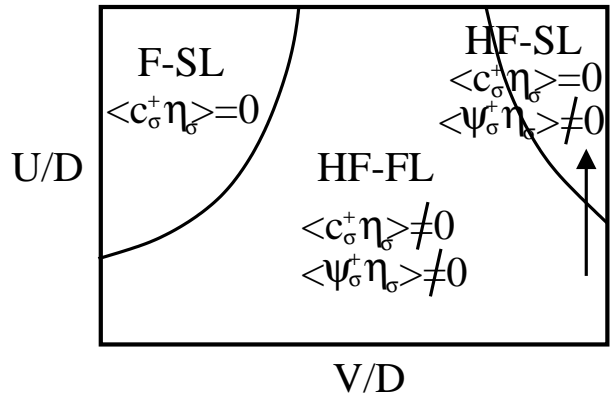}
\caption{Schematic phase diagram of the finite-U Anderson lattice
model without antiferromagnetism at zero temperature: HF-FL
(heavy-fermion Fermi liquid), HF-SL (heavy-fermion spin liquid),
and F-SL (fractionalized Fermi liquid)} \label{Fig. 1}
\end{figure}

We proposed the variational ground-state wave function for the
finite-U ALM since the stability of the heavy-fermion spin liquid
state against compact gauge fluctuations cannot be fully confirmed
in the effective field theory approach, thus it is necessary to
check its existence from the microscopic model. It will be
interesting to observe such a spin liquid state in the strong
hybridization and large repulsion limits, where the heavy-fermion
band still exists, but this corresponds to an insulator owing to
charge gap.

We close this paper discussing the possibility of
valance-fluctuation-induced superconductivity.\cite{Miyake}
Recently, we have generalized the U(1) slave-rotor formulation of
the Hubbard model into the SU(2) one, allowing not only local
density fluctuations but also pairing excitations.\cite{Kim_Rotor}
Both collective charge fluctuations form an SU(2) slave-rotor
matrix field, where its off-diagonal components are associated
with superconductivity. We believe that the SU(2) slave-rotor
decomposition is also available to the finite-U ALM. This
superconducting mechanism may explain physics in $PuCoGa_5$, where
superconductivity occurs with HF physics at the same
time.\cite{Coleman_Review,PuCoGa5} This interesting possibility is
under investigation.


\begin{thebibliography}{9}
\bibitem{LGW_Why} One cautious person may claim that the
slave-particle framework is not appropriate for the
Landau-Ginzburg-Wilson forbidden quantum transition in the
heavy-fermion context since no local gauge-invariant Kondo order
parameter exists, and no symmetry is broken in the Fermi-liquid
ground state while the Fermi liquid description in the
slave-particle context is in the condensation of slave-bosons,
thus breaking the internal U(1) gauge symmetry inevitably. This
statement is only partially correct. This situation reminds us of
the characterization of superconductivity. Usually, we interpret
superconductivity as the condensed phase of Higgs bosons (Cooper
pairs). This identification necessarily breaks the U(1) gauge
symmetry, thus cannot be correct in a rigorous manner. However, we
should admit that this characterization is useful. The way to
understand this identification is to break the gauge symmetry
explicitly, fixing one gauge. Since the gauge symmetry is
explicitly broken via gauge fixing, one can define the local order
parameter such as Cooper pair bosons. This is a physically
appealing way to understand such a boson-condensed phase in a
simple mean-field manner.
\bibitem{Rotor_model} Consider the boson-Hubbard model $H =
- t \sum_{\ij}(b_{i}^{\dagger}b_{j} + H.c.) +
\frac{U}{2}\sum_{i}n_{i}^{2}$ with $n_{i} = b_{i}^{\dagger}b_{i}$.
Using $b_{i} = \sqrt{n}e^{i\theta_{i}}$ and integrating out the
$n$ field, one can obtain the Lagrangian of this Hamiltonian, $L =
\frac{1}{2U}\sum_{i}(\partial_{\tau}\theta_{i})^{2} -
2tn\sum_{\ij}\cos(\theta_{i} - \theta_{j})$, the same as
$L_{\theta}$ of Eq. (8) in the mean-field approximation. It is
clear that this boson Hubbard model shows the QPT from the Mott
insulating phase to the superfluid state at $D/U \sim 1$.
\bibitem{TK_TFL} One need not be surprised at $T_{K} \sim V$
(instead of the exponentially small $T_{K}$) in the strong
hybridization limit because this is a usual result of the
mean-field treatment, considering the antiferromagnetic case
(Hartree-Fock calculation) where the magnetization order parameter
is exponentially small in the small-U limit of the Hubbard model
while it is proportional to Hubbard-U in the large-U limit. In the
strong hybridization approach the coherence temperature $T_{FL}$
would be exponentially small, corresponding to the
boson-condensation temperature. On the other hand, in the
conventional heavy-fermion metal-to-metal transition discussed in
section III, the Kondo temperature $T_{K}$ will be exponentially
small since this corresponds to the weak hybridization approach,
where $V_{eff} = |\langle e^{-i\theta}\rangle| V$ is an effective
hybridization coupling constant with the boson-condensation
amplitude $|\langle e^{-i\theta}\rangle|$. Actually, in the
single-impurity Anderson model this was indeed found in the
non-crossing approximation scheme.\cite{FG}
\end{thebibliography}
\end{document}